\documentclass[11pt,a4paper]{article}
\usepackage{amsmath,amssymb}
\usepackage{epsfig,graphicx}

\begin{document}

\title{\bf SELECTED PROBLEMS OF BARYONS SPECTROSCOPY: 
CHIRAL SOLITON VERSUS QUARK MODELS\footnote{Based partly on the talks presented at the
Workshop on Nuclear and Particle Physics at J-PARC (NP08), Mito, Ibaraki, 
Japan, March 05-07, 2008;
15th International Seminar on High Energy Physics Quarks-2008, Sergiev Posad, Russia, 
May 23-29 and at the International
Workshop on Hadron Structure
and QCD (HSQCD'2008), Gatchina, Russia, June 30 - July 4, 2008.}}
\author{Vladimir ~B.~Kopeliovich$$\footnote{{\bf e-mail}: kopelio@inr.ru},
\\
$$ \small{\em Institute for Nuclear Research of RAS} \\
\small{\em 60-th October Anniversary Prospect 7a, Moscow 117312}}
\maketitle

\begin{abstract}
Inconsistency between rigid rotator and bound state models at arbitrary
number of colors, rigid rotator --- soft
rotator dilemma and some other problems of baryon spectroscopy are discussed
in the framework of the chiral soliton approach (CSA). 
Consequences of the comparison of CSA results with simple quark models are considered
and the $1/N_c$ expansion for the effective strange antiquark mass is presented, as it follows
from the CSA.
Strong dependence of the effective strange antiquark mass on the SU(3) multiplet is 
required to fit the CSA predictions. The difference of `good' and `bad'  diquark masses,
which is about 100 Mev, is in reasonable agreement with other estimates. Multibaryons
(hypernuclei) with strangeness are described and some states of interest are 
predicted within CSA as well.
\end{abstract}

\section{Introduction}
 In spite of (or due to?) recent dramatic events with (non)observation of narrow pentaquark states,
the studies of baryons spectrum --- 
nonstrange, strange, and with heavy flavors --- remain to be very actual for accelerator physics.
Discovery of baryon states besides well established (e.g., octet, decuplet and certain resonances)
could help to the progress in understanding hadrons structure.

In the absence of the complete theory of strong interactions there are different approaches
and models of hadron structure; each has some advantages and certain drawbacks.
Interpretation of hadrons spectra in terms of quark models (QM) is widely
accepted, 
QM are ``most successful tool for the classification and interpretation'' (R.Jaffe) of hadrons 
spectrum. These models are so widely accepted because (probably) they correspond to our intuitive 
ideas how the bigger object --- baryon, for example --- can be made of smaller ones --- quarks.
However, our intuition, based on macroscopic experience, may be totally misleading in the world 
of elementary particles.

QM are to large extent phenomenological
since there are no regular methods of solving relativistic many-body problem.
The number of constituents (e.g. additional $q\bar q$-pairs) and their weight should not be fixed 
as starting condition, in a true relativistic theory, but should be obtained by means of solving
adequite relativistic equations (and the quark confinement should be obtained in this way, as well!).

In view of this global unresolved problem alternative approaches are of interest. In particular, the 
chiral soliton approach (CSA) 
based on few principles represented by the model lagrangian, 
has certain advantages.
Baryons and baryonic systems are considered on equal footing (the look `from outside').
CSA has many features of a true theory, but still it is a model:
some elements of phenomenology are present necessarily in CSA as well.
Results obtained within CSA mimic some features of baryons spectrum within quark models
due to Gell-Mann --- Okubo relations for the masses of baryons within definite $SU(3)$ multiplet.
\section{Features of the chiral soliton approach}
 The CSA is based on fundamental 
principles and ingredients incorporated in the {\it truncated effective chiral lagrangian}:
$$L^{eff} = -{F_\pi^2\over 16}Tr l_\mu l_\mu + {1\over 32e^2}Tr [l_\mu l_\nu]^2
+{F_\pi^2m_\pi^2\over 8}Tr \big(U+U^\dagger -2\big)+..., \eqno(1) $$
the chiral derivative $l_\mu = \partial_\mu U U^\dagger,$ $U\in SU(2)$ or $\in SU(3)$-
unitary matrix depending on chiral fields, 
$m_\pi$ is pion mass, $F_\pi$-pion decay constant taken from experiment, $e$ - the only 
parameter of the model which defines the weight of the antisymmetric 4-th order in chiral derivatives
term in the lagrangian (Skyrme term)
\footnote{In some papers the constant $F_\pi$ and even the mass $m_\pi$
are considered as parameters, although they are fixed by existing data. Such approach is useful, 
however, 
for investigations of some global properties of chiral soliton models and multiskyrmions.}.
The effective lagrangian (1) can be deduced from underlying lagrangian of Quantum Chromodynamics
\cite{de}, and in this way the infinite number of terms appear. The higher order in $l_\mu$ 
terms are not shown in (1). 6-th order term is taken into account in a 
number of calculations, and it does not change the properties of multiskyrmions considerably.
The mass term $\sim F_\pi^2m_\pi^2$,
 changes asymptotics of the profile $f$ and the structure
of multiskyrmions at large $B$. For the $SU(2)$ case
$$ U= cos f + i\, (\vec{n}\vec{\tau}) sin\,f, \eqno(2)$$
unit vector $\vec{n}$ depends on 2 functions, $\alpha,\;\beta$.
Three profiles $\{f,\;\alpha,\;\beta\}(x,y,z)$ parametrize the 
unit vector on the 3-sphere $S^3$.

The soliton is configuration of chiral fields, 
possessing topological charge identified with the baryon number $B$ (Skyrme, 1961):
$$ B= {-1\over 2\pi^2}\int s_f^2 s_\alpha I\left[(f,\alpha,\beta)/(x,y,z)\right] d^3r \eqno(3)$$
where $I$ is the Jacobian of the coordinates transformation, $s_f=sin\,f,\; s_\alpha =sin \alpha$,
etc. 
So, the quantity $B$ shows how many times $S^3$ is covered when 
integration over $R^3$ is made.
Recall that surface of the unit sphere $S^3$ equals
$$\int s_f^2s_\alpha df d\alpha d\beta=2\pi^2.\eqno(4)$$

Minimization of the classical mass functional $M_{cl}$ for each value of baryon number 
provides 3 profiles $f,\alpha,\,\beta$, the mass of static configuration and
allows to calculate  binding energies of classical configurations, moments of 
inertia $\Theta_I,\;\Theta_J,\;\Theta_K$ and some other
characteristics of chiral solitons wich contain implicitly  
information about interaction between baryons and are necessary to perform the quantization
procedure --- to get the spectrum of baryon states with definite quantum numbers.

\section{Skyrmions quantization and spectrum of baryons}
The observed spectrum of states is obtained by means of quantization procedure 
and depends on quantum numbers of baryons and above mentioned properties of classical 
configurations, moments of inertia, $\Sigma$-term ($\Gamma$), etc.
In $SU(2)$ case the rigid 
rotator model (RRM)\cite{anw} is most effective and successfull in describing the
properties of nucleons, $\Delta$-isobar, some properties  light nuclei
(N.Manton, S.Krusch and S.Wood, \cite{mmw}) and also the so called ``symmetry energy'' 
of nuclei with $A \lesssim 20$ \cite{ksm}.

In the $SU(3)$ case different quantization models have been developed. Probably, mostly accepted 
way to get the spectrum of baryons is to place the
established $SU(2)$ classical configuration (e.g. the so called `hedgehog' for the $B=1$ skyrmion) 
in the upper left corner of the $SU(3)$ matrix of chiral fields and to quantize the $SU(3)$ 
zero modes corresponding to rotations in the $SU(3)$ configuration space \cite{gua}. 
The following mass formula takes place, corresponding to this rigid rotator model:
$$ M(p,q,Y,I,J) = M_{cl} + {K(p,q,I_R)\over 2\Theta_K} + {J(J+1)\over 2\Theta_\pi} +
\delta M_{(p,q)}(Y,I), \eqno(5)$$
$$\qquad \qquad \qquad  \sim N_c \qquad \sim 1 \qquad \quad \sim N_c^{-1} \qquad \quad \sim 1,$$
which is in fact expansion in powers of $1/N_c$.
$$ K(p,q,I_R) = C_2(SU3) - I_R(I_R+1) -N_c^2B^2/12,\quad C_2(SU3) =(p^2+q^2+pq)/3 +p+q,  $$
$p,\,q$ are the numbers of upper and lower indeces in the spinor describing the $SU(3)$ multiplet,
$I_R$ is the so called `right' isospin, $I_R=J$ - the value of spin of the $B=1$ state.
Some paradox is in the fact that total splitting of the whole multiplet is $\sim N_c$.

Mass splittings $\delta M$ are due to the term in the lagrangian
$$ {\cal L}_M \simeq - \tilde m_K^2 \Gamma {s_\nu^2\over 2}, \eqno(6)$$
$\nu$ is the angle of rotation into `strange' direction, 
$\tilde m_K^2 = F_K^2m_K^2/F_\pi^2 -m_\pi^2$ includes $SU(3)$-symmetry violation in
flavor decay constants. For accepted values of the model parameters numerical values of
some important characteristics of the $B=1$ skyrmion are: $\Gamma \sim 5\, Gev^{-1} \sim \Sigma $,
moments of inertia $\Theta_\pi \sim (5 - 6)Gev^{-1},\;\Theta_K \sim (2 - 3)Gev^{-1}.$
All inertia are proportional to the number of colors, $\Theta \sim N_c$.

The multiplets of exotic baryons are shown in Fig. 1. Recall that for the `octet' 
$[p,q]=[1,(N_c-1)/2]$,
for `decuplet' $[p,q]=[3,(N_c-3)/2]$, and $(p+2q) = N_c$. For exotic multiplets shown in Fig.1
$p+2q = N_c+3$ \footnote{This particular choice of $[p,q]$ values is really a result of convention
for large $N_c$ generalization of the model. For this choice the upper states within each 
$SU(3)$ multiplet at arbitrary $N_c$ coincide with those at $N_c=3$.}.
The lower index in notations of states indicates the isospin of the state, e.g.
$$ \Phi/\Xi_{3/2}= |\overline{10},S=-2,I=3/2>,\;\Sigma_2 = |27,S=-1,I=2>, 
\;\Omega_1 = |27,S=-3,I=1>. $$

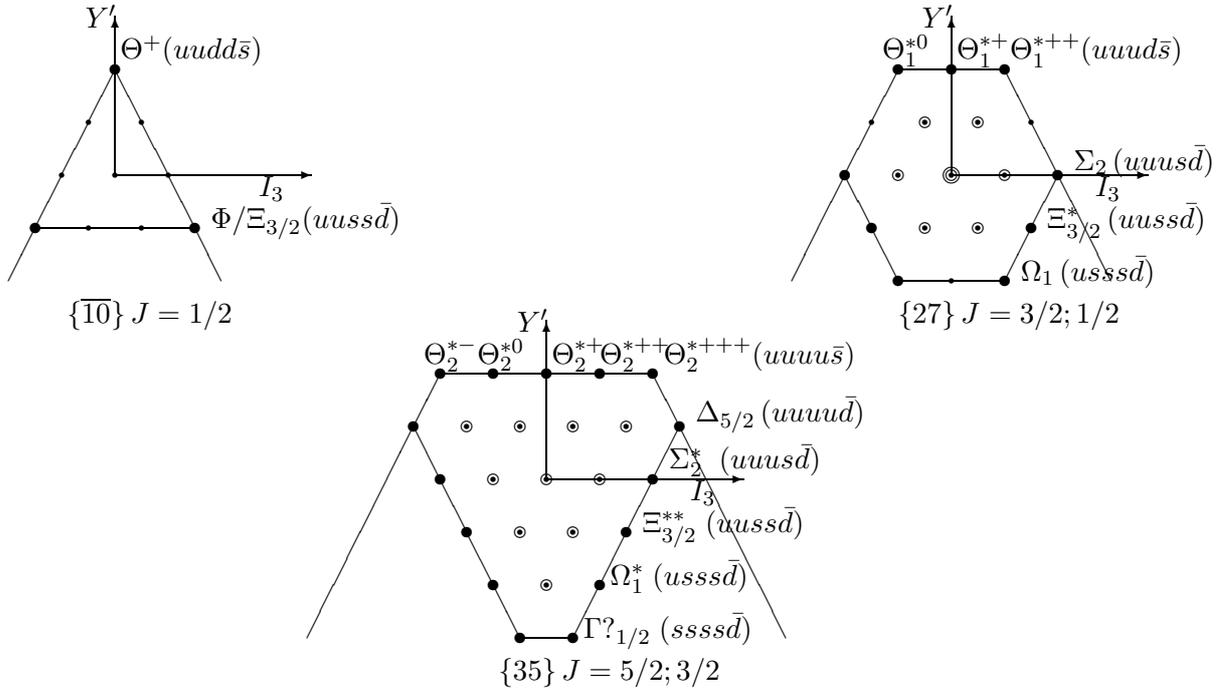
\begin{figure}
\label{multiplet}
\setlength{\unitlength}{.7cm}
\newcommand\Axis{%
   \begin{picture}(1,1)
   \put(0,0){\vector(1,0){3.7}}
   \put(0,0){\vector(0,1){3.}}
   \put(-0.55,2.8){$Y'$}
   \put(2.7,-0.4){$I_3$}
   \end{picture}}

\begin{center}
\begin{picture}(5,5) 
\put(3,4){\circle*{0.18}}
\put(1.5,1){\circle*{0.18}}
\put(4.5,1){\circle*{0.18}}

\multiput(2.5,3)(1,0){2}{\circle*{0.1}}
\multiput(2,2)  (1,0){3}{\circle*{0.1}}
\multiput(2.5,1)(1,0){2}{\circle*{0.1}}

\put(3,4){\line(-1,-2){2.}}
\put(3,4){\line(1,-2){2.}}
\put(1.5,1){\line(1,0){3}}

\put(3,2){\Axis}
\put(2.1,-0.8){$\{\overline{10}\}\, J=1/2$}
\put(3.1,4.2){$\Theta^+ (uudd\bar{s})$}
\put(4.8,1.0){$\Phi/\Xi_{3/2} (uuss\bar{d})$}
\end{picture} 
\hfill
\begin{picture}(5,5) 
\multiput(0,4)(1,0)    {3}{\circle*{0.18}}
\multiput(-1,2)(0.5,-1) {3}{\circle*{0.18}}
\multiput(3,2)(-0.5,-1){3}{\circle*{0.18}}

\multiput(-0.5,3)(1,0){4}{\circle*{0.1}}
\multiput(0,2)  (1,0){3}{\circle*{0.1}}
\multiput(.5,1)(1,0){2}{\circle*{0.1}}
\multiput(1,0)  (1,0){1}{\circle*{0.1}}

\multiput(.5,3)(1,0){2}{\circle{0.2}}
\multiput(0,2)  (1,0){3}{\circle{0.2}}
\multiput(.5,1)(1,0){2}{\circle{0.2}}
\put(1,2){\circle{0.3}}

\put(0,4){\line(1,0){2}}
\put(0,4){\line(-1,-2){2}}
\put(4,0){\line(-1,2){2}}
\put(3,2){\line(-1,-2){1}}
\put(0,0){\line(1,0){2}}
\put(0,0){\line(-1,2){1}}

\put(1,2){\Axis}
\put(0,-0.8){$\{27\}\, J=3/2;1/2$}
\put(-0.3,4.2){$\Theta^{*0}_1$}
\put(1.1,4.2){$\Theta^{*+}_1$}
\put(2.1,4.2){$\Theta^{*++}_1(uuud\bar{s})$}
\put(3.3,2.1){$\Sigma_2\,(uuus\bar{d})$}
\put(2.8,1.0){$\Xi^*_{3/2}\,(uuss\bar{d})$}
\put(2.3,0.0){$\Omega_1\,(usss\bar{d})$}
\end{picture} 
\\[.5cm]
\begin{picture}(5.5,6) 
\multiput(0.5,5)(1,0)    {5}{\circle*{0.18}}
\multiput(0,4)  (0.5,-1) {5}{\circle*{0.18}}
\multiput(5,4)  (-0.5,-1){5}{\circle*{0.18}}

\multiput(1,4)  (1,0){4}{\circle*{0.1}}
\multiput(1.5,3)(1,0){3}{\circle*{0.1}}
\multiput(2,2)  (1,0){2}{\circle*{0.1}}
\multiput(2.5,1)(1,0){1}{\circle*{0.1}}

\multiput(1,4)  (1,0){4}{\circle{0.2}}
\multiput(1.5,3)(1,0){3}{\circle{0.2}}
\multiput(2,2)  (1,0){2}{\circle{0.2}}
\multiput(2.5,1)(1,0){1}{\circle{0.2}}

\put(0.5,5){\line(1,0){4}}
\put(0,4){\line(1,-2){2}}
\put(5,4){\line(-1,-2){2}}
\put(2,0){\line(1,0){1}}
\put(-2,0.){\line(1,2){2.5}}
\put(7,0.){\line(-1,2){2.5}}

\put(2.5,3){\Axis}
\put(1.6,-0.8){$\{35\}\, J=5/2;3/2$}
\put(0.2,5.2){$\Theta^{*-}_2$}
\put(1.2,5.2){$\Theta^{*0}_2$}
\put(2.6,5.2){$\Theta^{*+}_2$}
\put(3.5,5.2){$\Theta^{*++}_2$}
\put(4.7,5.2){$\Theta^{*+++}_2 (uuuu\bar{s})$}
\put(5.3,4.1){$\Delta_{5/2}\,(uuuu\bar{d})$}
\put(4.8,3.2){$\Sigma_2^*\;\;(uuus\bar{d})$}
\put(4.3,2.0){$\Xi_{3/2}^{**}\;(uuss\bar{d})$}
\put(3.7,1.0){$\Omega_1^{*}\;(usss\bar{d})$}
\put(3.2,0.0){$\Gamma ?_{1/2}\;(ssss\bar{d})$}
\end{picture}
\end{center}

\vspace{.3cm}
\protect\caption{The $I_3-Y'$ diagrams $(Y'=S+1)$ for multiplets of pentaquark baryons, antidecuplet,
$[p,q]=[0,(N_c+3)/2]$, $\{27\}$-plet with $[p,q]=[2,(N_c+1)/2]$ and $\{35\}$-plet, $[p,q]=[4,(N_c-1)/2]$. 
For $N_c>3$ these diagrams should be extended within long
lines, as shown in the picture. Quark contents are given for manifestly exotic states 
(components with maximal value of $I_3$), when $N_c=3$.}
\end{figure}

``Strangeness contents'' 
$$ C_S = <s_\nu^2/2>_B  \eqno(7)$$
can be calculated exactly with the help of wave functions
in $SU(3)$ configuration space, for arbitrary number of colors $N_c$ \cite{vkpen,ksrev}.

Some examples of values of $C_S$ at arbitrary number of colors $N_c$ are presented in
Tables 1 and 2 taken from \cite{ksrev} \footnote{For 
the case of `nucleon' strangeness content at arbitrary $N_c$ has been presented first in \cite{kk}.}.

{\newcommand\rbox[1]{\makebox[66pt][r]{$#1$}}
\begin{center}
\begin{tabular}{|l|l|l|l|l|l|}
\hline
$[p,q];\;(Y', \,I)$& $\qquad  \qquad \qquad C_S(N)$ & $C_S(N=3)$ \\
\hline
$[1,(N-1)/2]$& &\\
\hline
$\;Y'=1,\,I=1/2$  & $\rbox{2(N+4)}/[(N+3)(N+7)]\quad$& $\;7/ 30$\\
\hline
$\;Y'=0,\,I=0$  & $\rbox3/(N+7)$ &$\;9/ 30$\\
$\;Y'=0,I=1$  & $\rbox{(3N+13)}/[(N+3)(N+7)]$& $11/ 30$\\
\hline
$*Y'=-1,I=3/2$ & $\rbox{(4N+18)}/[(N+3)(N+7)]$& --- \\
$\;Y'=-1,I=1/2$ & $\rbox4/(N+7)$&$ 12/ 30$\\
\hline
$[3,(N-3)/2]$& &\\
\hline
$\;Y'=1,\,I=3/2$  & $\rbox{2(N+4)}/[(N+1)(N+9)]$& $\;7/ 24$\\
\hline
$\;Y'=0,\,I=1$  & $\rbox{(3N+7)}/[(N+1)(N+9)]$ &$\;8/ 24$\\
\hline
$*Y'=-1,\,I=5/2$ & $\rbox{(4N+22)}/[(N+1)(N+9)]$& --- \\
$\;Y'=-1,\,I=1/2$ & $\rbox{(4N+6)}/[(N+1)(N+9)]$& $\;9/ 24$\\
\hline
$*Y'=-2,I=3$ & $\rbox{(5N+29)}/[(N+1)(N+9)]$& --- \\
$\;Y'=-2,I=0$ & $\rbox5/(N+9)$&$10/24$ \\
\hline
$[0,(N+3)/2]$& &\\
\hline
$\;Y'=2,\,I=0$ & $\rbox3/(N+9)$& $\;6/ 24$\\
$\;Y'=1,\,I=1/2$ & $\rbox{(4N+9)}/[(N+3)(N+9)]$ &$\;7/ 24$\\
$\;Y'=0,\,I=1$ & $\rbox{(5N+9)}/[(N+3)(N+9)]$& $\;8/ 24$\\
$\;Y'=-1,\,I=3/2$& $\rbox{(6N+9)}/[(N+3)(N+9)]$&$ \;9/ 24$\\
$*Y'=-2,\,I= 2$& $\rbox{(7N+9)}/[(N+3)(N+9)]$& --- \\
\hline
\end{tabular}
\end{center} }
{\bf Table 1.} {Strangeness contents of the "octet", "decuplet" and "antidecuplet" of
baryons at arbitrary $N=N_c$, for unmixed states, $Y'=S+1$. Few states (marked by $*$) are shown 
which appear only if $N>3$, mostly they are states with maximal possible value of isospin at
fixed $Y'$.}

Approximately at large $N_c$
$$C_S \simeq {2+|S|\over N_c}. \eqno(8)$$
The Gell-Mann --- Okubo formula takes place in the form
$$  C_S = -A(p,q)\,Y - B(p,q) \left[ Y^2/4 - \vec I^2\right] +C(p,q), \eqno(9)$$
$ A_(p,q), B_(p,q), C_(p,q)$ depend on particular $SU(3)$ multiplet.
For the `octet', for example \cite{ksrev},
$$ A("8") = {N_c+2 \over (N_c+3)(N_c+7)},\quad B("8") = {2 \over (N_c+3)(N_c+7)},\quad
 C("8") = {3 \over (N_c+7)}.\eqno(10)$$

\newcommand\rbox[1]{\makebox[106pt][r]{$#1$}}
\begin{center}
\begin{tabular}{|l|l|l|l|l|l|}
\hline
$[p,q];\;(Y',\,I)$&$\qquad  \qquad \qquad C_S(N)$& $C_S(N=3)$\\
\hline
$[2,(N+1)/2]$& & \\
\hline
$\;Y'=2,\,I=1$ & $\rbox{(3N+23)}/[(N+5)(N+11)]$& $32/ 112$\\
\hline
$\;Y'=1,\,I=3/2$ & $\rbox{(4N^2+65N/2-3/2)}/[(N+1)(N+5)(N+11)]$& $33/ 112$\\
$\;Y'=1,\,I=1/2$ & $\rbox{(4N+24)}/[(N+5)(N+11)]$& $36/ 112$\\
\hline
$\;Y'=0,\,I=2$ & $\rbox{(5N^2+39N-26)}/[(N+1)(N+5)(N+11)]$ & $34/ 112$\\
$\;Y'=0,\,I=1$ & $\rbox{(5N^2+33N+8)}/[(N+1)(N+5)(N+11)]$ & $38/ 112$\\

$\;Y'=0,\,I=0$ & $\rbox{5}/(N+11)$ & $5/ 14$\\
\hline
$*Y'=-1,\,I=5/2$ & $\rbox{(6N^2+{91\over 2}N-{101\over 2})}/[(N+1)(N+5)(N+11)]$ & --- \\
$\;Y'=-1,\,I=3/2$ & $\rbox{(6N^2+38N-8)}/[(N+1)(N+5)(N+11)]$ & $40 / 112$\\
$\;Y'=-1,\,I=1/2$ & $\rbox{(6N+7/2)}/[(N+1)(N+11)]$ & $43/112$\\
\hline
$*Y'=-2,\,I=3$& $\rbox{(7N^2+52N-75)}/[(N+1)(N+5)(N+11)]$& ---\\
$\;Y'=-2,\,I=1$& $\rbox{(7N+2)}/[(N+1)(N+11)]$& $46/ 112$\\
\hline
$[4,(N-1)/2]$& &\\
\hline
$\;Y'=2,\,I=2$ & $\rbox{(3N+25)}/[(N+3)(N+13)]$& $34/96$\\
\hline
$\;Y'=1,\,I=5/2$ & $\rbox{(4N^2+85N/3-79)}/[(N-1)(N+3)(N+13)]$ & $21/96$\\
$\;Y'=1,\,I=3/2$ & $\rbox{(4N+24)}/[(N+3)(N+13)]$ & $36/96$\\
\hline
$*Y'=0,\,I=3$ & $\rbox{(5N^2+{104\over 3}N-133)}/[(N-1)(N+3)(N+13)]$ & ---\\
$\;Y'=0,\,I=2$   & $\rbox{(5N^2+{74\over 3}N-67)}/[(N-1)(N+3)(N+13)]$ & $26/96$\\
$\;Y'=0,\,I=1$& $\rbox{(5N+23)}/[(N+3)(N+13)]$ & $38/96$\\
\hline
$*Y'=-1,\,I=7/2$& $\rbox{(6N^2+41N-187)}/[(N-1)(N+3)(N+13)]$ & ---\\
$\;Y'=-1,\,I=3/2$& $\rbox{(6N^2+21N-55)}/[(N-1)(N+3)(N+13)]$ & $31/96$\\
$\;Y'=-1,\,I=1/2$& $\rbox{(6N+22)}/[(N+3)(N+13)]$ &$40/96$ \\
\hline
$*Y'=-2,\,I=4$& $\rbox{(7N^2+{142\over 3}N-241)}/[(N-1)(N+3)(N+13)]$ & ---\\
$\;Y'=-2,\,I=1$& $\rbox{(7N^2+52N/3-43)}/[(N-1)(N+3)(N+13)]$ & $36/96$\\
$\;Y'=-2,\,I=0$& $\rbox{7}/(N+13)$ & $42/96$\\
\hline
$*Y'=-3,\,I=9/2$& $\rbox{(8N^2+{161\over 3}N-295)}/[(N-1)(N+3)(N+13)]$&---\\
$\;Y'=-3,\,I=1/2$& $\rbox{(8N-31/3)}/[(N-1)(N+13)]$             & $41/96$\\
\hline
\end{tabular}
\end{center}
 {\bf Table 2.} {Strangeness contents for unmixed states of the $"\{27\}"$-plet
 (spin $J=3/2$) and
 $"\{35\}"$-plet ($J=5/2$) of baryons, for arbitrary $N=N_c$ and  numerically for $N_c=3$.
 As in Table 1, some states which exist only for $N_c>3$ (with maximal isospin)
 are marked with $*$.}\\

For `decuplet'
$$ A("10") = {N_c+2 \over (N_c+1)(N_c+9)},\quad B("10") = {2 \over (N_c+1)(N_c+9)},\quad
 C("10") = {3 \over (N_c+9)},\eqno(11)$$
and for `antidecuplet' where for each iso-multiplet a relation takes place $I = (1-S)/2$ 
it was possible to obtain the relations
$$ A("\overline{10}")+{3\over 2}B("\overline{10}")  = {N_c \over (N_c+3)(N_c+9)},\quad
 C("10")- 2B("\overline{10}") = {5N_c+9 \over (N_c+3)(N_c+9)}.\eqno(12)$$

If we try to make expansion in $1/N_c$, then parameter is $\sim 7/N_c$ - for the `octet'.
For `decuplet' and `antidecuplet' expansion parameter is $\sim 9/N_c$ and becomes worse
for greater multiplets, $"\{27\}"$-plet, $"\{35\}"$-plet, etc.
Apparently, for realistic world with $N_C=3$ the $1/N_c$ expansion {\it does not work}.

Any chain of states connected by relation $ I= C' \pm Y/2$ reveals
linear dependence on hypercharge (strangeness). Interpretation of these 
results in terms of strange quark/antiquark masses should be quite careful.
For such multiplets as `octet', `decuplet' the CSA {\it mimics 
the quark model} with effective strange quark mass
$$m_s^{eff} \sim \tilde m_K^2 \Gamma\, \bigl[A(p,q)\mp 3 B(p,q)/2\bigr]. \eqno(13)$$
This is valid if the flavor symmetry breaking (FSB) is included 
in the lowest order of perturbation theory. At large $N_c$
$$ m_s^{eff} \sim \tilde m_K^2 \Gamma/N_c,\eqno(14)$$
too much, about $\sim (0.6\,-\,0.7) \,GeV$ if extrapolated to $N_c=3$.

If we make expansion in RRM, we obtain  for the `octet' of baryons the contribution to the
mass, proportional to $\bar m_K^2$
$$\delta M_N =2 \tilde m_K^2 {\Gamma\over N_c} \Biggl(1 - {6\over N_c} \Biggr),\quad
\delta M_\Lambda = \bar m_K^2 {\Gamma\over N_c}\Biggl(3 - {21\over N_c} \Biggr), $$
$$\delta M_\Sigma = \bar m_K^2 {\Gamma \over N_c}\Biggl(3 -{17\over N_c} \Biggr),\quad
\delta M_\Xi = \tilde m_K^2 {\Gamma \over N_c}\Biggl(4 - {28\over N_c} \Biggr), \eqno(15)$$
and for decuplet
$$\delta M_\Delta = 2\tilde m_K^2 {\Gamma\over N_c} \Biggl(1 - {6\over N_c} \Biggr),...\quad
\delta M_\Omega = \bar m_K^2 {\Gamma\over N_c}\Biggl(5 - {45\over N_c} \Biggr), \eqno (15a) $$
equidistantly for all 4 components.
Note, that for the `nucleon' and `$\Delta$' these contributions to the mass coincide
in the leading and next-to-leading orders of $1/N_c$ expansion, and can be considered
as contribution of the `sea' of $s\bar s$ pairs. The effective strange quark masses estimates
and their $1/N_c$ expansion follow from (15) immediately (section 6).
\section{Bound state model of skyrmions quantization}
Within the bound state model (BSM) \cite{ck}
anti-kaon or kaon is bound by $SU(2)$ skyrmion. The mass formula takes place
$$ M = M_{cl} + \omega_S + \omega_{\bar S} + |S| \omega_S + \Delta M_{HFS},\eqno(16)$$
where flavor and antiflavor excitation energies 
$$\omega_S= N_c(\mu-1)/8\Theta_K \simeq {\bar m_K^2\Gamma\over N_c},\;\;
\omega_{\bar S}= N_c(\mu+1)/8\Theta_K \simeq {N_c \over 4\Theta_K} +{\bar m_K^2\Gamma\over N_c},
\eqno(17)$$
$$\mu = \sqrt{1+\bar m_K^2/M_0^2}\simeq 1+{\bar m_K^2\over 2M_0^2} 
= 1+8{\bar m_K^2\Gamma\Theta_K\over N_c^2},$$ 
$$ \omega_S + \omega_{\bar S} = {\mu N_c\over 4\Theta_K}\simeq {N_c\over 4\Theta_K} + 
{\bar m_K^2N_c\over 8\Theta_K M_0^2}.  $$
$$ M_0^2=N_c^2/(16\Gamma\Theta_K)\sim N_c^0,\quad \mu \sim N_c^0 \sim 1. \eqno(18)$$ 
The expansion of $\mu$ made above really does not work well even for the case of
strangeness, however, it is very useful for comparison of BSM and RRM.

The hyperfine splitting correction depending on hyperfine splitting
constants $c$ and $\bar c$, isospin $I$ and spin $J$ of the state, and ``strange isospin'' 
$I_S=|S|/2$ equals \cite{ck}
$$\Delta M_{HFS} = {J(J+1)\over 2\Theta_\pi} 
+\frac{(c_S-1)[J(J+1)-I(I+1)] +(\bar c_S-c_S)
I_S(I_S+1)}{2\Theta_\pi}.\eqno(19)$$
$$ c_S =1 - {\Theta_\pi \over 2\mu\Theta_K} (\mu-1) \simeq 1- 4{\Theta_\pi \Gamma m_K^2\over N_c^2}, $$
$$ \bar c_S = {\Theta_\pi \over \mu^2\Theta_K} (\mu-1) \simeq 1-8{\Theta_\pi 
\Gamma m_K^2\over N_c^2}. \eqno(20)$$
The approximate equality shown in the right side takes place when expansion in $m_K^2$ is possible. 
In this approximation $\bar c_S \simeq c_S^2$, as mentioned in the literature. 
It is a point of principle that in the BSM baryon states are labeled by their strangeness (flavor), 
spin and isospin, but do not belong apriori to definite $SU(3)$ multiplet $(p,q)$. They can be a 
mixture of different $SU(3)$ multiplets, indeed.

For flavor (negative strangeness 
or beauty, positive charm) the HFS correction dissappears if $\bar m_K=0$, and we can rewrite
the mass formula for flavored states:
$$M(I,J,S)\simeq M_{cl} + {N_c \over 4\Theta_K} +{J(J+1)\over 2\Theta_\pi} +$$
$$+{\bar m_K^2 \Gamma\over N_c}
\left\{2+|S| -{2\over N_c}\left[J(J+1)-I(I+1)+I_S(I_S+1)\right]\right\}. \eqno (21)$$
It is clear from this expression that the energy is minimal when `strange' isospin is maximal,
i.e. $I_S=-S/2$. For decuplet isospin $I=(3+S)/2$ and $I_S(I_S+1)-I(I+1) = -5(3+2S)/4$, therefore,
 equidistant location of the components of decuplet is reproduced.

In this way we obtain for the `octet' and `decuplet' the contributions depending on $m_K^2$:
$$\delta M_N =2 \tilde m_K^2 {\Gamma\over N_c}, \quad  
\delta M_\Lambda = \tilde m_K^2 {\Gamma \over N_c}\Biggl(3 - {3\over N_c} \Biggr), $$
$$\delta M_\Sigma = \tilde m_K^2 {\Gamma \over N_c}\Biggl(3 + {1\over N_c} \Biggr), \quad
\delta M_\Xi = \tilde m_K^2 {\Gamma \over N_c}\Biggl(4 - {4\over N_c} \Biggr), $$
$$\delta M_\Delta =2 \tilde m_K^2 {\Gamma\over N_c} \simeq \delta M_N, \quad  
\delta M_\Omega = \tilde m_K^2 {\Gamma \over N_c}\Biggl(5 - {15\over N_c} \Biggr), \eqno(22) $$

It is instructive to compare the total splitting of the `octet' and `decuplet' in the BSM and in RRM
$$ \Delta_{tot}(`8',BSM) = \tilde m_K^2 {\Gamma \over N_c}\Biggl(2 - {4\over N_c} \Biggr), \quad
 \Delta_{tot}(`8',RRM) = \tilde m_K^2 {\Gamma \over N_c}\Biggl(2 - {16\over N_c} \Biggr), $$
$$ \Delta_{tot}(`10',BSM) = \tilde m_K^2 {\Gamma \over N_c}\Biggl(3 - {15\over N_c} \Biggr), \quad
... \quad \Delta_{tot}(`10',RRM) = \tilde m_K^2 {\Gamma \over N_c}\Biggl(3 - {33\over N_c} \Biggr).
\eqno(23)$$    
In BSM mass splittings are bigger than in RRM.

It follows already from this comparison that the RRM used for prediction of pentaquarks 
\cite{pras} is {\it different} from  
the BSM model, used in \cite{itz} to disavow the $\Theta^+$
\footnote{Intensive discussion of the CSA predictions validity for exotic baryon states
has been initiated in \cite{coh}. However, the explicit difference between RRM and BSM
in the next to leading terms of $1/N_c$ expansion of contributions $\sim \bar m_K^2$,
being discussed here, was not established in \cite{coh}.}.

For antiflavor (positive strangeness or beauty, or negative charm) certain changes should be done:
$\omega_S \to \omega_{\bar S}$ and $c_S \to c_{\bar S}$ in the last term.
It is crucially important that for the antiflavor 
 the hyperfine splitting constants are different; they can be obtained by means of the 
change $\mu \to -\mu$ in above formulas (see, e.g. detailed evaluation in \cite{ksrev}):
$$c_{\bar S} = 1 -{\Theta_\pi \over 2\mu\Theta_K} (\mu+1) \simeq
1 - {\Theta_\pi \over \Theta_K} + 4{\Theta_\pi \Gamma m_K^2\over N_c^2} + O(m_K^4), $$
$$\bar c_{\bar S} = 1 +{\Theta_\pi \over \mu^2\Theta_K} (\mu+1) \simeq
1 +2 {\Theta_\pi \over \Theta_K} -24{\Theta_\pi \Gamma m_K^2\over N_c^2} + O(m_K^4), \eqno(24)$$
and even approximate inequality of the type $\bar c_S \simeq c_S^2$ does not hold for 
positive strangeness.

As a result, the mass formula for anti-flavored states takes the form:
$$M(I,J,S>0)\simeq M_{cl} + {N_c+S \over 4\Theta_K} +{J(J+1)\over 2\Theta_\pi} +{1\over 2\Theta_K}
\left[I(I+1)-J(J+1)+3I_S(I_S+1)\right] +$$
$$+{\bar m_K^2 \Gamma\over N_c}
\left\{2+|S| +{2\over N_c}\left[J(J+1)-I(I+1)-7I_S(I_S+1)\right]\right\}. \eqno (25)$$

For antiflavor (positive strangeness, etc.) the term in Eq. (25) $\sim 1/\Theta_K$
is large even for small $m_K^2$:
$$ \Delta M_{HFS}^{\bar S} (\bar m_K=0) = {J(J+1)\over 2\Theta_\pi} +{1\over 2\Theta_K}
\left\{[-J(J+1) + I(I+1)] +3I_S(I_S+1)\right\}. \eqno (18a) $$
This contribution to the position of the baryon mass agrees with the result of the
RR model. 

The case of exotic $S=+1$ $\Theta$ hyperons is especially interesting. 
For $\Theta_0^+ \in \overline{10}$,  $J=1/2, \;I=0$, and we obtain
$$M_{\Theta_0,J=1/2}= M_{cl}+ {2N_c+3\over 4\Theta_K} + {3\over 8\Theta_\pi} + 
\bar m_K^2\Gamma\Biggl({3\over N_c}-{9\over N_c^2}\Biggr),$$
For $\Theta_1^+ \in \{27\}$,  $J=3/2, \;I=1$, and we have
$$M_{\Theta_1,J=3/2}= M_{cl}+ {2N_c+1\over 4\Theta_K} + {15\over 8\Theta_\pi} + 
\bar m_K^2\Gamma\Biggl({3\over N_c}-{7\over N_c^2}\Biggr),$$
For $\Theta_0^+ \in \{35\}$ $J=5/2, \;I=2$, and the contribution to the mass is
$$M_{\Theta_2,J=5/2}= M_{cl}+ {2N_c-1\over 4\Theta_K} + {35\over 8\Theta_\pi} + 
\bar m_K^2\Gamma\Biggl({3\over N_c}-{5\over N_c^2}\Biggr). \eqno(26)$$

The terms $\sim 1/\Theta_K$ agree with those obtained in RRM 
for anti-decuplet, $\{27\}$- and $\{35\}$-plets (terms, proportional to $K(p,q,J)$ in the
RRM mass formula).
This means that, indeed, we can interprete these positive strangeness states as belonging to definite
$SU(3)$ multiplets --- antidecuplet, $\{27\}$- and $\{35\}$=plets \footnote{It is an unresolved
 problem, however, how to obtain other components of these multiplets within the BSM. Evaluations 
performed in the literature are not sufficient for this purpose. The point is that, for example,
strange isospin which is unique for the states with strangeness $S=\pm 1$, is uncertain for the 
components of exotic multiplets different from the $S=1$ states \cite{ksrev}.}, at least when 
expansion of the quantity $\mu$, we made above. is possible.

We should compare also the contributions $\sim \bar m_K^2\Gamma$ with the mass splitting 
correction from RRM:
$$\delta M^{RRM}_{\Theta_0,J=1/2}=\bar m_K^2\Gamma\Biggl({3\over N_c}-{27\over N_c^2}\Biggr), \quad
\delta M^{RRM}_{\Theta_1,J=3/2}=\bar m_K^2\Gamma\Biggl({3\over N_c}-{25\over N_c^2}\Biggr),$$
$$\delta M^{RRM}_{\Theta_2,J=5/2}=\bar m_K^2\Gamma\Biggl({3\over N_c}-{23\over N_c^2}\Biggr),\eqno(27)$$
and again --- as in case of `octet' and `decuplet' --- considerable difference takes place
between the RRM and BSM results.

The addition of the term to the BSM result,  {\it possible due to normal ordering 
ambiguity for the operators of (anti)strangeness production, 
present in BSM} (I.Klebanov, VBK, 2005, unpublished)
$$ \Delta M_{BSM-RRM} = -6\bar m_K^2 {\Gamma\over N_c^2}(2+|S|) \eqno(28)$$
brings results of RRM and BSM in agreement --- for nonexotic as well as exotic $S=+1$ states.
This procedure looks however not quite satisfactorily: if we believe to RRM, why we need BSM at all?
Anyway, RRM and BSM in its accepted form are {\it different models}.

The rotation-vibration approach (RVA) by H.Weigel and H.Walliser \cite{ww} tries to
unify RRM and BSM in some way,  $\Theta^+$ has been confirmed with somewhat higher
energy and considerable width $(\Gamma_\Theta \sim 50\,MeV)$\footnote{The alternative 
RRM --- BSM is not resolved appropriately in the literature. In some cases when
there is an ambiguity, the priority is given to the RR model (see, e.g. \cite{ww},
sections 3 and 4). The hyperfine splitting correction in \cite{ww} has the form different from our.
According to Eq. (3.21) of \cite{ww} it is
$$\Delta M_{S} = {1\over 2\Theta_\pi} 
+\left[c_SJ(J+1)+(1-c_S)I(I+1) +c_S(c_S- 1)/4\right],\eqno(3.21)$$ the last term being
completely different from our in Eq. (19). In view of this, the authors of \cite{ww} stated:
"The comparison with the RRA suggests that these quartic terms contribute $9/8\Theta_K$ to
the mass of the $S=1$ baryons." According to our BSM formulas we have 
$(\bar c_{\bar S} - c_{\bar S})I_S(I_S+1)/(2\Theta_\pi)|_{\bar m_K^2=0}
 =9/(8\Theta_K)$ in agreement with the RRM,
and there is no need to correct the BSM formulas `by hands'.}.

\section{The role of configuration mixing}
Configuration mixing due to the term $ \sim m_K^2 \Gamma s^2_\nu$ in the lagrangian \cite{ya} is 
important, e.g. the $\Delta$ state from decuplet of baryons is 
mixed with the $\Delta'$ state from $\{27\}$-plet, and as a result, the splitting between 
these states becomes larger: the mass of the $\Delta$ goes down, and the mass of 
$\Delta'$ increases, see Fig.2. Similar mixing takes place for other baryon states which have equal values
of strangeness and isospin but belong to different $SU(3)$ multiplets.

For anti-decuplet mixing decreases slightly the total splitting,
and pushes the $N^*$ and $\Sigma^*$ states toward higher energy. Mixing with components of the octet
is important. Apparent contradiction with the simplest assumption of equality of masses
of strange quarks and antiquarks $m(s) = m(\bar s)$ takes place (see next section).

For decuplet mixing increases total splitting considerably, but
approximate {\bf equidistancy still remains}!\footnote{Therefore, statement made in several 
papers that approximate equidistancy within the decuplet of baryons is an argument that configuration 
mixing is negligible, is not correct}. Mixing with the components of $\{27\}$-plet
is important, e.g. the $\Delta \in\{10\}$ after mixing with $\Delta^* \in\{27\}$ moves to 
the lower value of mass. 

 A note for the QM should be made: states with different numbers 
of $q\bar q$ pairs can mix, and such mixing
{\it should be taken into account}. In the diquark-diquark-antiquark picture proposed in \cite{jw}
the mixing of pentaquark states with the ground state baryon octet should be included,
since strong interactions do not conserve the number of quark-antiquark pairs present in the hadron.
This mixing pushes the pentaquark states towards higher energy and changes the whole picture
of relative positions of baryon states. Without such mixing the diquark picture \cite{jw} looks 
artificial, whereas within CSA this problem is resolved in natural way.

We conclude this section with the following discussion of the case of large value of the mass
$\bar m_F$, which, besides $m_K$, can be also $\bar m_D$ or $\bar m_B$. When this mass is large enough,
the expansion of the quantity $\mu$ in (17) cannot be made, and  instead of this expansion
we have $\mu \simeq \bar m_F/M_0  = 4\bar m_F \sqrt{\Gamma\Theta_K}/ N_c.$ This linear dependence of
$\mu$ and also flavor excitation energies $\omega_F,\,\bar \omega_F$  on the mass 
$m_F$, given by (17), is quite reasonable, but
\begin{center}
{\bf
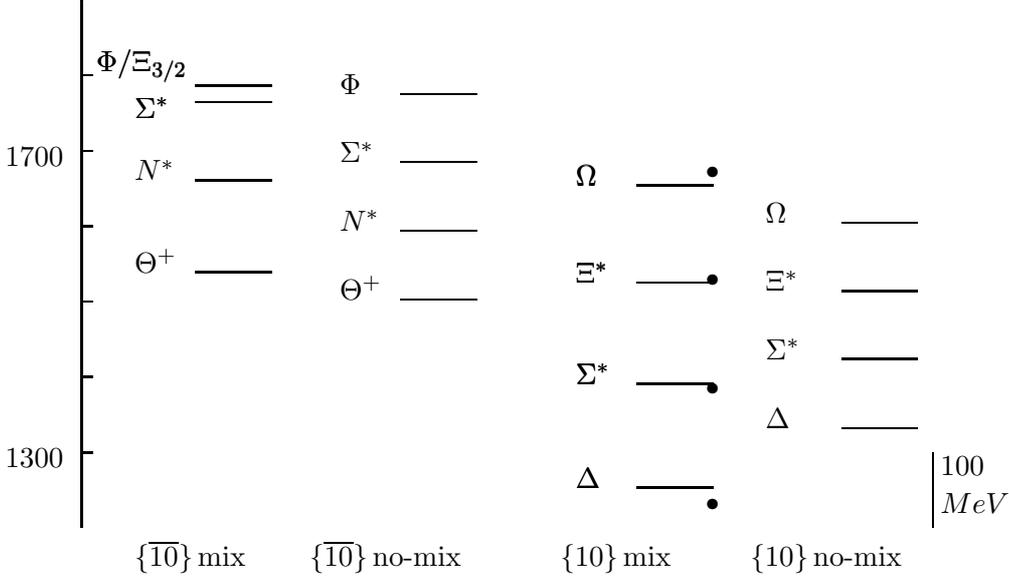
\begin{figure}[h]
\label{mix-influence}
\setlength{\unitlength}{1.cm}
\begin{flushleft}
\begin{picture}(13,6.5)

\put(0.5,0){\line(0,1){7.}}
\put(0.5,1){\line(1,0){0.15}}
\put(0.5,2){\line(1,0){0.15}}
\put(0.5,3){\line(1,0){0.15}}
\put(0.5,4){\line(1,0){0.15}}
\put(0.5,5){\line(1,0){0.15}}
\put(0.5,6){\line(1,0){0.15}}
\put(0.5,8){\line(1,0){0.15}}
\put(-.5,0.8){$1300$}
\put(-.5,4.8){$1700$}

\put(11.7,0){\line(0,1){1.}}
\put(11.8,0.7){$100$}
\put(11.8,0.2){$MeV$}

\put(2,5.86){\line(1,0){1.}}
\put(2,5.868){\line(1,0){1.}}
\put(2,5.864){\line(1,0){1.}}
\put(2,5.64){\line(1,0){1}}
\put(2,5.644){\line(1,0){1}}
\put(2,4.61){\line(1,0){1}}
\put(2,4.614){\line(1,0){1}}
\put(2,3.39){\line(1,0){1}}
\put(2,3.394){\line(1,0){1}}

\put(0.7,6.064){\bf $\Phi/\Xi_{3/2}$}
\put(0.7,6.06){\bf $\Phi/\Xi_{3/2}$}
\put(1.205,5.44){\bf $\Sigma^*$}
\put(1.2,5.44){\bf $\Sigma^*$}
\put(1.2,4.61){{\bf $N^*$}}
\put(1.2,3.39){{\bf $\Theta^+$}}

\put(1.2,-0.5){$\{\overline {10}\}\, $mix}

\put(4.7,5.75){\line(1,0){1.}}
\put(4.7,4.85){\line(1,0){1}}
\put(4.7,3.94){\line(1,0){1}}
\put(4.7,3.03){\line(1,0){1}}

\put(3.9,5.75){\bf $\Phi$}
\put(3.9,4.85){\bf $\Sigma^*$}
\put(3.9,3.94){{\bf $N^*$}}
\put(3.9,3.03){{\bf $\Theta^+$}}

\put(3.5,-0.5){$\{\overline {10}\}\, $no-mix}

\put(7.8,4.544){\line(1,0){1.}}
\put(7.8,4.54){\line(1,0){1.}}
\put(7.8,3.254){\line(1,0){1.}}
\put(7.8,3.25){\line(1,0){1.}}
\put(7.8,1.914){\line(1,0){1.}}
\put(7.8,1.91){\line(1,0){1.}}

\put(7.8,0.534){\line(1,0){1.}}
\put(7.8,0.53){\line(1,0){1.}}

\put(8.8,4.72){\circle*{0.15}}
\put(8.8,3.3){\circle*{0.15}}
\put(8.8,1.85){\circle*{0.15}}
\put(8.8,0.32){\circle*{0.15}}

\put(7.,4.544){\bf $\Omega$}
\put(7.,4.54){\bf $\Omega$}
\put(7.,3.254){\bf $\Xi^*$}
\put(7.,3.25){\bf $\Xi^*$}
\put(7.,1.914){\bf $\Sigma^*$}
\put(7.,1.91){\bf $\Sigma^*$}
\put(7.,0.534){\bf $\Delta$}
\put(7.,0.53){\bf $\Delta$}

\put(6.8,-0.5){$\{10\}\, $mix}

\put(10.5,4.044){\line(1,0){1.}}
\put(10.5,3.142){\line(1,0){1.}}
\put(10.5,2.239){\line(1,0){1.}}
\put(10.5,1.325){\line(1,0){1.}}

\put(9.5,4.044){\bf $\Omega$}
\put(9.5,3.142){\bf $\Xi^*$}
\put(9.5,2.239){\bf $\Sigma^*$}
\put(9.5,1.325){\bf $\Delta$}

\put(9.3,-0.5){$\{10\}\, $no-mix}

\end{picture}
\end{flushleft}
\vspace{0.5cm}
\protect\caption{{\it Influence of the configuration mixing} (H.Yabu, K.Ando, 1988 \cite{ya}) on the 
mass splitting within antidecuplet and decuplet of baryons, RR model 
(variant by H.Walliser \cite{wk}).
For decuplet the data are shown by black points.}
\end{figure}}
\end{center}
 it is not possible to ascribe the quantized states
uniquely to definite $SU(3)$ irreps, as we made above in section 4. It is a challenging problem 
to get such linear in $m_F$ behaviour of flavored 
states energies within rotator models, RRM or SRM. Probably, strong configuration mixing
which should take place in this case, would be able to reduce the quadratic dependence on $m_F$
(or linear in $\Gamma$) and to convert it to linear dependence. Numerical calculations with 
configuration mixing program
arranged by H.Walliser and used in \cite{wk} confirm this point, but an 
analytical proof is desirable.
\section{Comparison of CSA results with simple quark model}
 It is possible to make comparison of the CSA results with expectations
from simple quark model in {\it pentaquark} approximation (projection of CSM on QM).
The masses $m_s$, $m_{\bar s}$ and the mass of $s\bar s$ pair $m(s\bar s)$ come into play,
as presented in Table 3 for pure states (without mixing).
Examples of wave functions of pentaquarks (PQ-s) in the diquark-diquark-antiquark picture by
R.Jaffe and F.Wilczek \cite{jw} are the following (see also \cite{clo}, \cite{vkpen}.):
$$\Theta_0 \in \{\overline{10}\}\sim [ud][ud]\bar s,$$ 
where $[ud]$ is diquark with zero isospin 
(singlet in flavor).
Other states can be obtained, e.g. by action of the operator $U^-$ which transforms $d$-quark 
into $s$-quark, and $\bar s \to \bar d$, $U^-\bar s = -\bar d$, and well known 
isotopic $I^{\pm}$ operators. For example, 
$$N^{*+}\in \{\overline{10}\}\sim [\sqrt 2\bar s \{[us][ud]\} - \bar d [ud][ud]]/\sqrt 3,\, ...$$
$$ \Phi/\Xi_{3/2}^{--}\in \{\overline{10}\}\sim [sd][sd]\bar u,\, ...
... \Phi/\Xi_{3/2}^{+}\in \{\overline{10}\}\sim [su][su]\bar d. $$
For larger $N_c$ the number of diquarks,
equal to $N_D=(N_c+1)/2$, increases, and additional $s\bar s$ pairs appear in wave functions of
some states \footnote{Standard assumption is that the baryon number of the quark equals to $1/N_c$.
We accept also the relation between hypercharge and strangeness in the form $Y=S + N_cB/3$ (see, e.g.
\cite{coh}). Note that the quantity $Y'$ in Fig.1 and Tables 1,2 is by definition $Y'=S+1$. The
wave function of `pentaquark' in this case is $"\Theta_0" \in "\{\overline{10}\}"
\sim [ud]...[ud]\bar s,$ with $(N_c+1)/2$ $[ud]$ diquarks, etc.}.

For antidecuplet at arbitrary $N_c$ according to Fig.1 and Table 1 any state with
strangeness $S$ has isospin $I =(1-S)/2$ and its mass equals
$$ M(\overline{10},S, I=(1-S)/2) = M(\overline{10},S=1,I=0) + \bar m_K^2\Gamma 
{(1-S)N_c \over (N_c+3)(N_c+9)}, \eqno (29)$$
Interpretation of this relation in terms of the quark model is not straightforward.
Simple relations can be obtained from Table 3 for the effective $s-$quark and
antiquark masses $m_s$ and $m_{\bar s}$, 
from the total splitting of antidecuplet $(N_c=3)$
$$[2m_s - m_{\bar s}]_{\overline{10}} = \bar m_K^2{\Gamma\over 8},\eqno(30)$$
which equals numerically $272\,MeV$  for the parameters accepted in \cite{wk}. 
For arbitrary number of colors this relation should be rewritten as
$$[(N_c+1)m_s -2 m_{\bar s}]_{\overline{10}} = \bar m_K^2{N_c\Gamma\over N_c+9}.\eqno(30a)$$
Configuration mixing
decreases this quantity to $247\,MeV $, see Table 3.
Relation (30) is the only relation which can be obtained, according to Table 3.
If we assume that the strange quark mass in antiecuplet is the same  as in decuplet,
$m_s(\overline{10}) = m_s(10)$, then the mass of strange antiquark should be negative
if the configuration mixing is not included: $m_{\bar s}(\overline{10}) = -m_s(\overline{10})$.
This relation looks unrealistic.
Note, that if the mass of $s$-antiquark within antidecuplet were equal to that of $s$-quark
(we call this variant {\it simplistic} model), 
then this splitting would be much smaller, just equal to $m_s \sim 130 \,Mev$. 
A natural way out of this contradiction is to allow the masses of strange quark/antiquark within
antidecuplet to be different from those within decuplet and other multiplets.

It is remarkable that
configuration mixing pushes the splitting towards simplistic quark model where the splitting
of antidecuplet should be about $m_s$, since $m_{\bar s} \simeq m_s$.
If we assume that the $s$-quark mass in $\overline{10}$ is about $150\,Mev$, as in the decuplet,
then strange antiquark within $\overline{10}$ should be very light, with the mass about $30 - 50\,Mev$.
\begin{center}
\begin{tabular}{|l|l|l|l|l|l|}
\hline
$|\overline{10},2,0>$&$|\overline{10},1,{1\over 2}>$&$|\overline{10},0,1>$&
$|\overline{10},-1,{3\over 2}>$& &\\
\hline
$m_{\bar s}$         &$2 m_{s\bar s}/3$   &$m_s+m_{s\bar s}/3$&$2 m_s$& & \\
\hline
564 & 655 & 745 & 836 & &\\
600 & 722 & 825 & 847 & & \\
\hline
\hline
$|27,2,1>$&$|27,1,{3\over 2}>$&$|27,0,2>$&$|27,-1,{3\over 2}>$&$|27,-2,1>$&\\
\hline
$m_{\bar s}$         &$m_{s\bar s}/2$   &$m_s$&$2 m_s$&$3m_s$  & \\

\hline
733 & 753 & 772 & 889 & 1005 &\\
749 & 887 & 779 & 911 & 1048 &\\
\hline
\hline
$|35,2,2>$&$|35,1,{5\over 2}>$&$|35,0,2>$&$|35,-1,{3\over 2}>$&$|35,-2,1>$&$|35,-3,{1\over 2}>$ \\
\hline
$m_{\bar s}$   &$0 $   &$m_s$&$2 m_s$&$3m_s$ &$4 m_s$ \\
\hline
1152 & 857 & 971 & 1084 & 1197 & 1311 \\
1122 & 853 & 979 & 1107 & 1236 & 1367 \\
\hline
\end{tabular}
\end{center}

{\bf Table 3.} Strange quark (antiquark) masses contributions (for the case $N_c=3$) 
and calculation results
within RRM, without and with configuration mixing (1-st and 2-d lines of numbers,
correspondingly \cite{wk}). For each value of strangeness the states with 
largest value of isospin are considered here.\\

For the components of $\{27\}$-plet with strangeness $S\leq -1$ the relation takes place
$$ M(27,S, I=(5+S)/2) = M(27,S=-1,I=2) - \bar m_K^2\Gamma 
{(S+1)\left(N_c^2-N_c+18\right) \over (N_c+1)(N_c+5)(N_c+11)}, \eqno(31)$$
and for $N_c=3$ we get $m_s^{eff} (27) \simeq 3\bar m_K^2\Gamma/56  \simeq  117\, Mev$.
which increases to $135\,MeV$ when configuration mixing is included.

From splittings within the $\{27\}$-plet between components with $Y' \geq 0$ we get also
$$[m_s - m_{\bar s}]_{27}= \bar m_K^2\Gamma/56,$$ 
which equals numerically to $39\,MeV$ \cite{wk} and reduces to  $30\,MeV$
when configuration mixing is included.

It is of interest that when configuration mixing is not included then the mass of 
strange quark-antiquark pair $m_{s\bar s}= (m_s + m_{\bar s})/2$ both for antidecuplet and
$\{27\}$-plet. This relation is the consequence of Gell-Mann --- Okubo relation, indeed.
For arbitrary $N_c$ interpretation of formula (31) in terms of effective quark/antiquark
masses becomes more difficult, since additional $s\bar s$ pairs are present in simple wave functions.

Let us consider now the biggest (in multiplicity) pentaquark. 
The remarkable property of $\{35\}$-plet is that the lowest in mass state is not the state
with highest value of hypercharge, $Y'=2$, but the state in the middle of the multiplet, which
has $Y'=1,\;S=0,\; I=5/2$. In the pentaquark approximation ($N_c=3$) this state contains 
neither strange quark/antiquark, no $s\bar s$ pair, and has the smallest numerically 
strangeness content among all baryons considered here. As can be seen from Table 3,
the mass of $s\bar s$ pair does not enter the masses of all
$\{35\}$-plet components with largest values of isospin.
The masses of these states with $S\leq 0$ are connected by relation
$$ M(35,S, I=5/2 +S/2) = M(35,S=0,I=5/2) - \bar m_K^2\Gamma 
{S\left(N_c^2+12 -11N_c/3\right) \over (N_c-1)(N_c+3)(N_c+13)}, \eqno(32)$$
so, for $N_c=3$ the quantity
$$ m_s^{eff} (35) = \bar m_K^2\Gamma {5 \over 96}\simeq 114 \,MeV \eqno(33) $$
can be considered as an effective strange quark mass in this case. Configuration mixing increases
this quantity to $130 \, MeV$ (see Table 3). 

From the difference between masses of $S=+1$ and $S=0$ states we can extract the effective strange
antiquark mass
$$[m_{\bar s}]_{35}= \bar m_K^2\Gamma {13 \over 96}\simeq 295\,MeV,\eqno(35)$$
which is remarkably large value. Configuration mixing reduces slightly this quantity to
$ 270\, MeV$.

For arbitrary $N_c$ we can get an idea about the behavior of strange antiquark mass for 
antidecuplet, $\{27\}$-plet and $\{35\}$-plet
if we make some assumption about the contribution of strange quark sea, in particular, that it is
the same as for `nucleon' and `$\Delta$'-isobar (coinciding in the leading and next-to-leading 
orders of $1/N_c$ expansion, see Table 1).

In this way we obtain from the RRM (the contribution of the sea
of $s\bar s$ pairs is subtracted):
$$ [m_{\bar s}]_{\overline{10}} \sim {\bar m_K^2 \Gamma \over N_c} \left(1 - {15\over N_c} \right);
\quad [m_{\bar s}]_{27} \sim {\bar m_K^2 \Gamma \over N_c} \left(1 - {13\over N_c} \right);
\quad [m_{\bar s}]_{35} \sim {\bar m_K^2 \Gamma \over N_c} \left(1 - {11\over N_c} \right), \eqno(36)$$
and within the BSM
$$ [m_{\bar s}]_{\overline{10}} \sim {\bar m_K^2 \Gamma \over N_c} \left(1 - {9\over N_c} \right);
\quad [m_{\bar s}]_{27} \sim {\bar m_K^2 \Gamma \over N_c} \left(1 - {7\over N_c} \right);
\quad [m_{\bar s}]_{35} \sim {\bar m_K^2 \Gamma \over N_c} \left(1 - {5\over N_c} \right).\eqno(37) $$
So, we see from here that numerical results shown in {\bf Table 3} can be understood qualitatively 
by this expansion, 
although extrapolation back to the real $N_c=3$ world is not possible. It is worth noting also
that the changes of the effective s-antiquark mass from antidecuplet to \{35\}-plet are equal within
RRM and BSM, although the mass itself is smaller in RRM --- in the next to leading order of
$1/N_c$ expansion.

We summarize our results for the first two terms of the $1/N_c$ expansion of the effective strange 
quark and antiquark masses in Table 4. `Octet' and `decuplet' of baryons do not contain valent
$s\bar s$ pairs, the mass difference between components is defined entirely by valent strange quarks.
The mass $m_s$ is defined as a half of the total splitting for the `octet' and 1/3 of the total 
splitting for `decuplet'.
\begin{center}
\begin{tabular}{|l|l|l|l|l|l|}
\hline
&$\quad \{8\}$&\quad $\{10\}$ &$\quad \{\overline{10}\}$ &$\quad \{27\} $& $\quad \{35\}$\\
\hline
$m_s^{RRM}$&$ 1-8/N_c$ &$1- 11/N_c$&$\quad  -  $&$\quad - $ &$\quad - $ \\
\hline
$m_s^{BSM}$&$ 1-2/N_c$ &$1-5/N_c$&$\quad  - $&$\quad  - $ &$\quad  - $   \\
\hline
$m_{\bar s}^{RRM}$&$\quad -$ &$\quad - $&$ 1- 15/N_c$&$1-13/N_c$ &$1-11/N_c$ \\
\hline
$m_{\bar s}^{BSM}$&$\quad  - $ &$\quad  -$&$ 1- 9/N_c$&$1-7/N_c$ &$1-5/N_c$ \\
\hline
\end{tabular}
\end{center}

{\bf Table 4.} First terms of the $1/N_c$ expansion for the effective strange quark 
and antiquark masses within different $SU(3)$ multiplets, in units $\bar m_K^2 \Gamma/N_c$.
Empty spaces are left in the cases of theoretical uncertainty. The assumption concerning strange 
quarks/antiquarks sea, described in the text, should be kept in mind.\\

Strong dependence of the $s$-antiquark mass on the multiplet is required when we project
the results of CSA on simple quark model:
is it artefact of CSA, or is physically  significant --- is not clear now. The influence of 
the configuration mixing on the contribution of $m_s, \;m_{\bar s},\; m_{s\bar s}$ to 
baryon states should be included in more detailed consideration.


\section{Diquarks mass difference estimates} 
Estimates of the diquark mass differences can be made roughly using results obtained from CSA.
As it was suggested by F.Wilczek \cite{w} the singlet in spin diquark
$[q_1q_2]$, which is antitriplet $\bar 3_F$ in flavor, is called `good' diquark $d_0$,  
the triplet in spin diquark $(q_1q_2)$, which is $6_F$ in flavor, is called `bad' diquark $d_1$.
Both `good' and `bad' diquarks are antitriplets in color.
As it was presented in previous section, the wave function for pentaquarks from antidecuplet 
can be written in terms of diquark wave functions \cite{jw}, \cite{clo} as:

$\Theta_0 \in \{\overline{10}\}\sim [ud][ud]\bar s$, ...
$\Phi/\Xi_{3/2}^{--}\in \{\overline{10}\}\sim [sd][sd]\bar u $ ... .

It is not possible to built $\{27\}$ and $\{35\}$-plets from `good' diquarks only,
`bad' diquarks are needed, as can be illustrated well by these examples of wave functions
of positive strangeness baryons:

$\Theta_1^0 \in \{27\} \sim (dd)[ud]\bar s,$ $\Theta_1^+ \in \{27\} \sim (ud)[ud]\bar s,$
$\Theta_1^{++} \in \{27\} \sim (uu)[ud]\bar s,$\\

$\Theta_2^- \in \{35\} \sim (dd)(dd)\bar s$, $\Theta_2^0 \in \{35\} \sim (ud)(dd)\bar s$, ...
$\Theta_2^{+++} \in \{35\} \sim (uu)(uu)\bar s$.\\

It seems to be natural to ascribe the difference of rotation energies
for different multiplets, given by the term $\sim K(p,q,I_R)$ in expression (5) to the 
difference of masses of `bad' and `good' diquarks.
Since `bad' diquark is heavier, this is obvious reason why $\Theta_1$
is heavier than $\Theta_0$, and $\Theta_2$ is more heavy.

From the difference of $\{27\}$-plet and antidecuplet masses
$$ M(d_1)-M(d_0) \sim {3\over 2\Theta_\pi} - {1\over 2\Theta_K} \sim 100\,MeV. \eqno(38)$$
from  $\{35\}$-plet and $\{27\}$-plet mass difference
$$ M(d_1)-M(d_0) \sim{5\over 2\Theta_\pi}-  {1\over 2\Theta_K} \sim 250\,MeV . \eqno(39)$$
Qualitatively this result seems to be OK, in agreement with previous estimates \cite{w} and, 
e.g., lattice calculations \cite{bab},
but this picture should be too naive.
In particular, interaction between diquarks may be important, which makes the 
$\Theta_{5/2}$ even more heavy.
\section{Rigid Rotator --- Soft Rotator dilemma}
The rigid rotator model is a limiting case of the rotator model when 
deformations of skyrmions during rotation in $SU(3)$ configuration space are totally neglected.
In the soft rotator model, opposite to rigid rotator, it is supposed
that soliton is deformed under influence of FSB forces: static energy minimization
is made at fixed value of $\nu$. Dependence on $\nu$ of static characterstics of
skyrmions is taken into account in the quantization procedure.

\begin{figure}[h]
\label{levelsRRSR}
\setlength{\unitlength}{.8cm}
\begin{flushleft}
\begin{picture}(12,8)

\put(0.5,0){\line(0,1){8.}}
\put(0.5,1){\line(1,0){0.15}}
\put(0.5,2){\line(1,0){0.15}}
\put(0.5,3){\line(1,0){0.15}}
\put(0.5,4){\line(1,0){0.15}}
\put(0.5,5){\line(1,0){0.15}}
\put(0.5,8){\line(1,0){0.15}}
\put(-.5,0.8){$1600$}
\put(-.5,4.8){$2000$}

\put(12,0){\line(0,1){1.}}
\put(12.1,0.7){$100$}
\put(12.1,0.2){$MeV$}
\put(2,2.87){\line(1,0){1.}}
\put(2,2.56){\line(1,0){1}}
\put(2,1.69){\line(1,0){1}}
\put(2,0.66){\line(1,0){1}}

\put(1.2,2.97){\bf $\Phi$}
\put(1.2,2.36){\bf $\Sigma^*$}
\put(1.2,1.69){{\bf $N^*$}}
\put(1.2,0.66){{\bf $\Theta^+$}}

\put(1.2,-0.5){$\{\overline {10}\}\, $ RR}

\put(4.8,6.18){\line(1,0){1.}}

\put(4.8,2.37){\line(1,0){1.}}

\put(4.,6.18){\bf $\Phi$}

\put(4.,2.37){\bf $\Theta^+$}

\put(4.2,-0.5){$\{\overline {10}\}\, $ SR}

\put(8,4.59){\line(1,0){1.}}
\put(8,3.42){\line(1,0){1.}}
\put(8,2.28){\line(1,0){1}}
\put(8,3.09){\line(1,0){1}}
\put(8,2.){\line(1,0){1}}

\put(7.,4.59){\bf $\Omega_{1}$}
\put(7.,3.42){\bf $\Phi^*$}
\put(7.,2.28){\bf $\Sigma_2$}
\put(7,2.89){\bf $\Delta^*$}
\put(7,1.8){\bf $\Theta_1$}

\put(7.2,-0.5){$\{27\}\, $ RR}

\put(11,7.7){\line(1,0){1.}}
\put(11,5.9){\line(1,0){1}}
\put(11,3.8){\line(1,0){1}}
\put(11,3.4){\line(1,0){1}}

\put(10.,7.7){\bf $\Omega_1$}
\put(10.,5.9){\bf $\Phi^*$}
\put(10,3.8){\bf $\Sigma_2$}
\put(10,3.3){\bf $\Theta_1$}

\put(10.2,-0.5){$\{27\}\, $ SR}

\end{picture}
\end{flushleft}
\vspace{.5cm}
\protect\caption{ Comparison of the rigid rotator (RR) and soft rotator (SR)
models predictions for the masses of exotic baryons, 
antidecuplet and $\{27\}$-plets. Not all states are shown for the SR
model. The code for SR model used here was arranged by B.Schwesinger, H.Weigel (1992)\cite{Schw}.}
\end{figure}
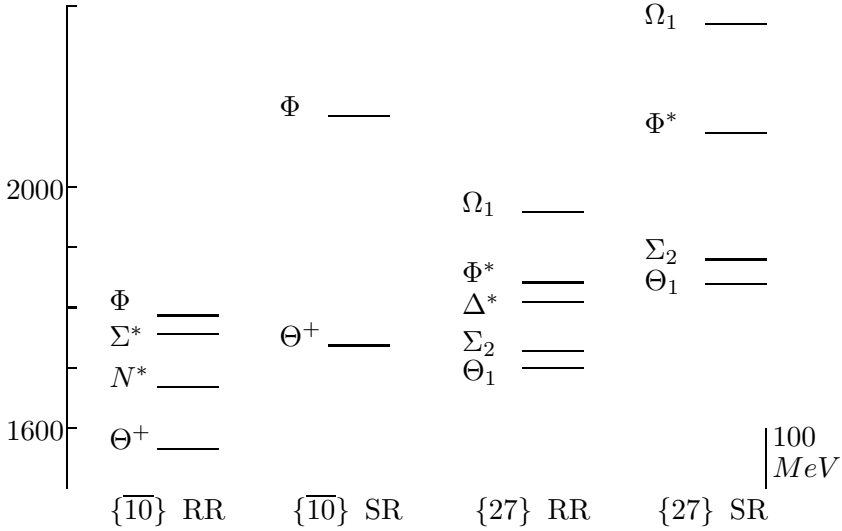
Static characteristics of skyrmions depend on $\nu$ - the angle of rotation into 
`strange' direction. For `strange', or kaonic inertia it is most important:
$$\Theta_K = {1\over 8}\int  (1-c_f)\Biggl[ F_K^2 - sin^2\nu (F_K^2-F_\pi^2)(2-c_f)/2+
+{1\over e^2}\biggl(f'^2+{2s_f^2\over r^2}\biggr) \Biggr] d^3r. \eqno(40)$$
It is decreasing function of $ sin^2\nu$. RRM corresponds to $\nu =0$, the maximal value of
kaonic inertia $\Theta_K $ and relatively low values of masses of exotic baryons (like
$\Theta, \; \Phi/\Xi_{3/2},$ etc.). Within SRM the masses of baryons from antidecuplet
and $\{27\}$ - plet are considerably greater than in RRM, mostly due to smaller value
of $\Theta_K $ (see Fig.3). The truth is somewhere between RR and SR models, but to make 
reasonable calculation seems to be unrealistic presently since the properties of baryonic
matter are not known, in particular the response of matter relative to the FSB forces.
\section{Strange multibaryons or hypernuclei}
Great advantage of CSA is that multibaryon states --- nuclei, hypernuclei ... --- 
can be considered on equal footing with the $ B=1$ case.
The rational map approximation proposed in \cite{hms}
 simplifies this work considerably and allows to calculate easily all static characteristics
of multiskyrmions necessary for spectrum evaluation. In particular, the $B$-number dependence
of quantities of interest has been established,
 $\Theta_I \sim B$, $\Theta_J \sim B^2$ for $B\leq 20 - 30$.
Some kind of the ``bag model'' for multibaryobs can be obtained with the help of this ansatz, 
starting with effective lagrangian \cite{vk}.
\def\br{\mbox{\boldmath $r$}}
\def\bm{\mbox{\boldmath $m$}}

\setlength{\unitlength}{1.cm}
\begin{flushleft}
\begin{figure}
\begin{picture}(12,6.)
\put(3,.5){\vector(1,0){2.5}}
\put(3,.5){\vector(0,1){3.8}}
\put(2.8,4.2){ $Y$}
\put(5.5,0.2){\bf $I_3$}
\put(0.5,4.7){ a) $Odd \;B\,,\; J=1/2, (3/2)$}
\put(2.3,3.2){$^3 H$}
\put(3.3,3.2){$^3 He$}
\put(2.8,2.3){$^3_\Lambda H$}
\put(2.5,3){\circle* {0.15}}
\put(3.5,3){\circle* {0.15}}
\put(2,2){\circle {0.15}}
\put(3,2){\circle* {0.15}}
\put(3,2){\circle {0.27}}
\put(4,2){\circle {0.15}}
\put(1.5,1){\circle {0.15}}
\put(2.5,1){\circle {0.15}}
\put(3.5,1){\circle {0.15}}
\put(4.5,1){\circle {0.15}}



\put(9,.5){\vector(1,0){2.3}}
\put(9,.5){\vector(0,1){3.8}}
\put(8.8,4.2){$Y$}
\put(11,0.2){$I_3$}
\put(6.5,4.7){ b) $Even \;B\,,\; J=0$}
\put(8.8,3.2){$^4 He $}
\put(8.3,2.3){$^4_\Lambda H$}
\put(9.3,2.3){$^4_\Lambda He$}

\put(9,3){\circle* {0.15}}

\put(8.5,2){\circle*{0.15}}
\put(8.5,2){\circle {0.27}}
\put(9.5,2){\circle*{0.15}}
\put(9.5,2){\circle {0.27}}

\put(8,1){\circle {0.15}}
\put(9,1){\circle {0.15}}
\put(10,1){\circle {0.15}}
\end{picture}
\vspace{-.2cm}
\protect\caption{ (a) The location of the isoscalar state (shown by double circle)
with odd $B$-number and $|S|=1$ in the upper part of the $(I_3 -Y)$ diagram.
(b) The same for isodoublet states with even $B$. The case of light hypernuclei
$_\Lambda H$ and $_\Lambda He$ is presented as an example. The lower parts of 
diagrams with $Y \leq B-3$ are not shown here.}
\end{figure}
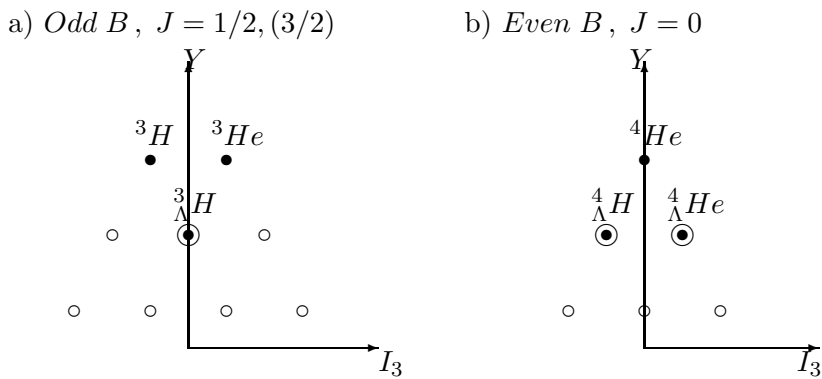
\end{flushleft}
Ordinary nuclei and hypernuclei (ground states) can be ascribed
to definite $SU(3)$ multiplets, as shown in Fig.4 for baryon numbers 3 and 4.
In a version of BSM it is possible to describe total binding energies
of light hypernuclei in qualitative, even semiquantitative agreement with data \cite{vk1}.
The collective motion of multiskyrmion in the SU(3) collective coordinates space 
is taken into account. The results of such estimates within the rigid oscillator model
(a variant of the bound state model) are presented in Fig.5, and quite satisfactory
qualitative agreement with existing data on total binding energies takes place.
\begin{figure}
\begin{center}
\epsfig{figure=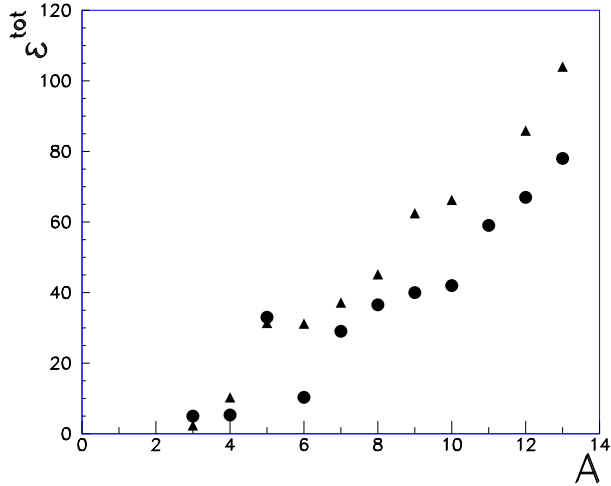,width=8.cm,angle=0}
\end{center}
\vspace{-.6cm}
\protect\caption{ Total binding energies of light hypernuclei (in MeV). Full triangles --- 
experimental data, full circles  --- theoretical results in a version of BSM applied to
multiskyrmions \cite{vk1}. Figure from \cite{ksm}b.}
\end{figure}

For the $B=2$ case more detailed investigations have been performed.
The lowest multiplets of dibaryons are shown in Fig.6: left figure shows the antidecuplet
of the $J=1$ dibaryons, the $I=0$ deuteron being the nonstrange state; right figure shows
the $J=0$  $\{27\}$-plet, the $I=1$ nucleon-nucleon scattering state being the upper 
(nonstrange) component.
There is also $\{35\}$-plet with the $N\Delta$-like nonstrange upper component (isospin $I=2$) and
$\{28\}$ - plet with the $\Delta\Delta$-like upper component (isospin $I=3$). $\{28\}$ - plet
contains the state with $S=-6$ (di-Omega). $\{35\}$-plet and $\{28\}$ - plet are not shown 
in Fig.6.

\vglue 0.1cm
\begin{figure}[h]
\label{multiplet}
\setlength{\unitlength}{1.cm}
\begin{flushleft}
\begin{picture}(12,6)
\put(3,2){\vector(1,0){2.5}}
\put(3,2){\vector(0,1){3}}
\put(2.6,4.7){$Y$}
\put(5.2,1.6){$I_3$}
\put(1,-1){$B=2,\;\{\overline {10}\},\, J=1 \,(2,...)$}
\put(3.1,4.1){$ D$}
\put(3.6,3.1){$ (\Lambda N)$}
\put(4.1,2.1){$ (\Xi N)$}
\put(4.6,1.1){$ (\Xi\Sigma)$}

\put(3,4){\circle*{0.21}}
\put(2.5,3){\circle {0.2}}
\put(3.5,3){\circle{0.2}}
\put(2,2){\circle{0.2}}
\put(3,2){\circle{0.2}}
\put(4,2){\circle{0.2}}
\put(1.5,1){\circle{0.2}}
\put(2.5,1){\circle{0.2}}
\put(3.5,1){\circle{0.2}}
\put(4.5,1){\circle{0.2}}

\put(1.5,1){\line(1,0){3}}
\put(1.5,1){\line(1,2){1.5}}
\put(4.5,1){\line(-1,2){1.5}}

\put(10,2){\vector(1,0){3}}
\put(10,2){\vector(0,1){3}}
\put(9.6,4.7){$Y$}
\put(12.7,1.6){$I_3$}
\put(8,-1){$B=2,\;\{27\},\, J=0\,(1,...)$}
\put(8.2,4.1){$ (nn)$}
\put(10.1,4.1){$(np)$}

\put(11.1,4.1){$(pp) $}
\put(11.6,3.1){$(\Sigma N) $}
\put(12.1,2.1){$(\Sigma\Sigma) $}
\put(11.6,1.1){$(\Sigma\Xi) $}
\put(11.1,.1){$(\Xi\Xi) $}

\put(9,4){\circle*{0.21}}
\put(10,4){\circle*{0.21}}
\put(11,2){\circle*{0.21}}

\put(8.5,3){\circle{0.2}}
\put(9.5,3){\circle*{0.1}}
\put(9.5,3){\circle {0.2}}
\put(10.5,3){\circle*{0.1}}
\put(10.5,3){\circle {0.2}}
\put(11.5,3){\circle{0.2}}

\put(8,2){\circle{0.2}}
\put(9,2){\circle*{0.1}}
\put(10,2){\circle*{0.1}}
\put(11,2){\circle*{0.1}}
\put(12,2){\circle{0.2}}
\put(9,2){\circle {0.2}}
\put(10,2){\circle {0.2}}
\put(11,2){\circle {0.2}}
\put(10,2){\circle {0.3}}

\put(8.5,1){\circle{0.2}}
\put(9.5,1){\circle*{0.1}}
\put(9.5,1){\circle {0.2}}
\put(10.5,1){\circle*{0.1}}
\put(10.5,1){\circle {0.2}}
\put(11.5,1){\circle{0.2}}

\put(9,0.){\circle{0.2}}
\put(10,0.){\circle{0.2}}
\put(11,0.){\circle{0.2}}

\put(8,2){\line(1,2){1}}
\put(8,2){\line(1,-2){1}}
\put(9,4){\line(1,0){2}}
\put(9,0){\line(1,0){2}}
\put(12,2){\line(-1,2){1}}
\put(12,2){\line(-1,-2){1}}

\end{picture}
\end{flushleft}
\vspace{.5cm}
\protect\caption{ $I_3-Y$ diagrams of multiplets of dibaryons, $B=2$: the $J=1$
antidecuplet (should not be mixed up with antidecuplet of pentaquarks, $B=1$) and
the $J=0$ $\{27\}$-plet. 
Virtual levels (scattering states) are shown in brackets, e.g. $(\Lambda N)$
scattering state which appears as near-threshold enhancement.}
\end{figure}

Calculations of spectrum of strange dibaryons have been performed in \cite{kss} in SRM which is more 
relevant for the $B=2$ case than for the $B=1$ case. When the $NN$-scattering state
was fitted to be on the right place (the deuteron binding energy is then about 30 Mev), 
all strange and multistrange dibaryons are
above threshold by few tens of Mev, so, they can appear as near-threshold enhancements in 
scattering cross sections of baryons with appropriate quantum numbers. These results are in 
qualitative agreement with quark models calculations \cite{gold}.

Multibaryons with positive strangeness or beauty (or negative charm) have been predicted within
similar approach as well \cite{ksj}.

Rotational excitations of any state have additional energy
$$\Delta E = {J(J+1)\over 2 \Theta_J}. \eqno(41) $$
$J=2^+$ excited states have energy by $\sim 2/\Theta_J$ greater than $\overline{10}$.
The state with $S=-1,\,I=1,\,J^P=2^+$ can be interpreted as $NN\bar K$ state with binding energy
$ \sim 100\;Mev$.
 For the $B=2 \;\;\{27\}$-plet $J=1$ states have energy by $1/\Theta_J\sim 60\, Mev$ 
greater than $J=0$ ground states.

The orbital inertia grows fast with increasing baryon (atomic) number,
$\Theta_J \sim B^p$, where $p$ is between 1 and 2. By this reason the number of rotational
states becomes larger for large baryon numbers.
Some of them can be interpreted as {\it deeply bound anti-kaon states } discussed 
intensively in \cite{aka} and other papers.
More detailed investigations of this issue are necessary.
\section{Summary and conclusions}
We can summarize our discussion in the following way:

The parameter of $1/N_c$ expansion is large for the case of the baryon spectrum, 
extrapolation to real world is not possible in this way, and conclusions made in the limit
$N_c\to \infty$ may not be valid in the real world.
Rigid (soft as well) rotator and bound state models coincide in
the first order of $1/N_c$ expansion, but {\it differ} in the next orders.

Configuration mixing is important, according to RRM, and makes substantial influence 
on the effective quark masses within simple quark model.

Transition to Soft Rotator Model from RRM may be crucial,
leading to the increase of masses, especially for exotic states.

 There is correspondence of the chiral soliton RRM and
 quark model predictions for pentaquarks spectra in negative $S$ sector
 of $\{27\}$ and $\{35\}$ plets: the effective mass of strange 
 quark is about $135 -130\,MeV$, slightly smaller for $\{35\}$.

 For positive strangeness components the link between 
CSM and QM requires strong dependence of effective $\bar{s}$ mass on 
particular $SU(3)$ multiplet. The $1/N_c$ expansion for the effective
strange antiquark mass provides different results within rotator and bound state models
in the next-to-leading order, but the changes of the effective $m_{\bar s}$ when we go 
from one multiplet to another
{are equal} for the RRM and BSM. Configuration mixing pushes spectra towards 
{\it simplistic} model - nice property, but reasons for this are not clear
presently.
 Diquarks mass difference estimates from CSA seem to be reasonable.

As conclusion, we state that chiral soliton models, based on few principles and
ingredients incorporated in effective lagrangian, allow to describe
qualitatively, in some cases even quantitatively, various chracteristics 
of baryons and nuclei --- from ordinary $(S=0)$ nuclei to known hypernuclei.
 This suggests that predictions of pentaquark states, as well as multibaryons 
with strangeness, are of interest. Existence of PQ by itself is without any doubt,
although very narrow PQ may not exist. Wide, even very wide PQ should exist, and
searches for PQ-s remain to be an actual task.

There are, however, problems when one tries to project results of
the CSA on quark models: strong dependence of strange antiquark mass on 
the $SU(3)$ multiplet; difference of masses of `bad' and `good' diquarks
is not unique in naive picture, at least.

In view of theoretical uncertainties, experimental investigations
are of crucial importance. In particular, experiments at the J-PARC 
accelerator (50 GeV) can provide 
 great chance to shed more light on the puzzles of baryon spectroscopy.

\section*{Acknowledgements} 
The author is indebted to H.Walliser and A.Shunderuk for fruitful collaboration. 
Helpful discussions with T.Cohen, 
R.Jaffe, I.Klebanov, N.Manton, C.Rebbi, H.Walliser and H.Weigel at different 
stages of the work are thankfully acknowledged, as well as useful remarks by R.Faustov.

\end{document}